\documentstyle[12pt,rotating,epsfig]{article}
\begin{document}
\begin{center}\Large
{\bf A Superheated Droplet Detector for Dark Matter Search}\\
\end{center}\normalsize
\begin{center}
L.A. Hamel, L. Lessard, L. Rainville and 
V. Zacek\footnote{Corresponding author. Tel. +514 343 6722 Fax. +514 343 6215 
E-mail zacekv@lps.umontreal.ca} \\
{\it Groupe de Physique des Particules, Universit\'e de Montr\'eal,
Canada}\\

and\\

Bhaskar Sur\\
{\it AECL, Chalk River Laboratories, Canada}\\
\end{center}
\vspace{3 cm}
\begin{abstract}
We discuss the operation principle of a detector based on
superheated droplets of Freon-12 and its feasibility for the search of weakly 
interacting cold dark 
matter particles. In particular we are interested in a neutralino search 
experiment in the mass range from 10 to 10$^{4}$ GeV/c$^{2}$ and with 
a sensitivity of 
better than 10$^{-2}$ events/kg/d. We show that our new proposed 
detector can be 
operated at ambient pressure and room temperature in a mode where it is 
exclusively sensitive to nuclear recoils like those following neutralino 
interactions, which allows a powerful background discrimination. An additional 
advantage
of this technique is due to the fact that the detection material, Freon-12, 
is cheap and
readily available in large quantities. Moreover we were able to show that 
piezoelectric 
transducers allow efficient event localization in 
large volumes.    
\end{abstract}
\newpage
\section {Introduction}

  Current models explaining the evolution of the universe and the measured 
slight
anisotropy of the cosmic background radiation have in common that they predict
an appreciable contribution of non-luminous, non-baryonic matter in the form of
a mixture of relativistic, light particles and non-relativistic, massive 
particles (so-called Hot and Cold Dark Matter). Accelerator experiments and 
results from the first round
of Dark Matter experiments explored up to now only a small range of masses and
types of possible candidate 
particles. In particular they leave room for an interpretation of 
Cold Dark Matter in terms of
weakly interacting massive particles with masses ranging from 20 GeV/c$^{2}$ 
up to a few hundred GeV/c$^{2}$. Here a
fitting candidate is the neutralino of the "minimal supersymmetric standard 
model" (\cite{bot}, \cite{griest}, \cite{mcdo}, \cite{good}). The cross section 
of these particles is expected to be 
of electroweak strength, with coherent or axial coupling. These particles 
are 
supposed to be concentrated in a spherical halo around our galaxy with a 
Maxwellian velocity distribution in the galactic frame with a mean velocity 
of about 300 km$/$s and a local density at the solar system of about 
0.3\, GeV/c$^{2}$/cm$^{3}$.
The detection reaction of neutralinos would be elastic scattering off a 
detector
nucleus and the nuclear recoil energy would be the measurable quantity. Under
the given assumptions the mean recoil energy would be\\
\begin{equation}
<E_{R}> \approx 2\,keV M_{N}\,(GeV/c^2)\; [M_{X}/(M_{N} + M_{X})]^2 
\end{equation}
where $M_{N}$ and $M_{X}$ are the masses of the detector nucleus and of the 
neutralino respectively.\\

For all detector materials the recoil energies are expected to be smaller 
than 100 keV and depending on more detailed assumptions on the cross section 
the expected event rates in the range between 4 to 20 keV are between 0.01 
to 100 events/kg/d. Therefore in order to ensure a reasonable count rate a 
large target mass of more than 50 to 100 kg 
is needed, especially if one wants to detect the slight annual variation in
count rate of several percents due to the relative motion of the earth around 
the sun. The latter 
would be the decisive signature for the detection of Dark Matter candidates.
Due to background limitations, however, current experimental 
sensitivities are still far too small ( $>$ factor 100) in order to reach 
the small interaction
rates predicted. The main reason for this is the fact that present detectors 
are sensitive to all kinds of ionizing 
radiations. Therefore an extremely low level of radioactive impurities in 
the detector 
material and its surroundings is necessary, as well as powerful active and
passive background rejection techniques.\\ 

 In order to reduce the overall 
background sensitivity we suggested in \cite{zac1} a detection method
which is exclusively sensitive to the high ionization density of recoiling 
nuclei with A $>$10, but which is insensitive to the much 
smaller specific ionization of $\alpha, \beta, \gamma$-radiation. 
As is well known from bubble
chamber operation, vapor droplet formation in superheated liquids is a process
of precisely this kind. But in contrast to the usual cyclic bubble chamber 
operation, the detection of nuclear recoils requires only moderately 
superheated liquids and a quasi-continuous operation becomes possible.
A further advantage of this technique is its operation at ambient 
temperatures, thus avoiding the need of cryogenics.\\
\begin{figure}[h]
\caption{Mean recoil energies expected in  Dark Matter search experiments with
detector liquids like $CCl_{2}F_{2}$ (Freon-12) and $CF_{3}Br$ as a function of 
neutralino mass $M_X$.}
\end{figure}
 In this paper we discuss systematic studies of an interesting technical 
realization of this 
detection principle which promises to be ideally suited 
for the detection of supersymmetric dark matter candidates, namely detectors 
based on superheated freon droplets (\cite{zac2},\cite{col}). After a general 
review of the detection principle, we investigate more in detail Freon-12 
based detectors and discuss 
systematic studies concerning the response of our newly developed counters with
3\% Freon loading to
neutron induced recoils, as well as to $\alpha$ - particles and recoiling 
nuclei in $^{241}$Am spiked detectors. Measuring the $\alpha$ - detection
efficiency and understanding the physical processes involved in the 
detection of these particles is of fundamental
importance, since we expect one of the main background contributions to 
the detection of dark matter particles to
be due to the recoils triggered by $\alpha$ - emitting activities in the 
decay chains of naturally present uranium and thorium contaminations.  
We describe systematic studies
of the detector response as a function of operating temperature and
present experimental tests on event detection and 
localization with piezolelectric 
sensors. We will show that modules with volumes as large as 22 l can be 
efficiently read out with a position resolution better than 0.5 cm. Finally  
the discussion of background sensitivity and overall efficiency will lead 
us to the conclusion that this new method has the potential of a substantially 
increased sensitivity for dark matter searches at relatively modest cost.\\ 

\section{Operation Principle} 
 It is generally 
agreed that the "heat spike" theory of Seitz describes well the process of 
vaporization of superheated liquids \cite{seitz}. In this model it is 
assumed that 
proto-bubbles are created by thermal spikes of released heat on a particle 
track. 
The growth of proto-bubbles is damped by adiabatic cooling, heat conduction 
and the viscosity of the liquid. Only if the seed cavity reaches a 
critical volume of radius $R_{c}$, it is large enough to nucleate a 
macroscopic liquid-to-vapor phase transition. $R_{c}$ is defined 
by $R_{c}$ = 2$\sigma (T)$
/$\delta p$, where $\sigma (T)$ is the surface tension of the 
liquid-vapor interface at 
the operating temperature T and $\delta p = p_{v}(T) -p_{0}$ is the difference 
between the vapor pressure in the bubble and the externally applied pressure.
Bubble formation is triggered if an amount of energy larger than a certain
critical energy $E_{c}$ is deposited within a region of extension 
$L \le 2R_{c}$. 
$E_{c}$ can be calculated and is well approximated for Freon-12 by

\begin{equation}
E_{c}(T) = (16\pi/3) (\sigma^{3}(T)/\delta p^{2}(T))
\end{equation}

$\delta p$ measures the degree of superheat and the larger $\delta p$, the 
smaller is $R_{c}$ and also the less heat is required for drop vaporization. 
However as
pointed out in \cite{wang}, \cite{harp}, \cite{riep} only a small fraction of 
the energy actually
deposited $E_{dep}$ by an ionizing particle within the thermal spike 
length $L \le 2R_{c}$ can be used to form a critical bubble. 
Therefore an  efficiency factor $\eta = E_{c}/E_{min}$  
determines the energy threshold $E_{min}$ for recoil detection and 
$E_{min}$ together with 
$L$ have to be determined 
experimentally. As will be shown below for Freon-12 the data suggest a value 
of $\eta$
between 2 - 6\%.\\

\section{Technical Realisation: The Superheated Droplet Detector}
  For our prototype studies we used superheated droplet 
detectors as they are used in neutron dosimetry (\cite{apf}, \cite{ing}). 
In these counters 
5 to 100$\mu$m diameter droplets of 
superheated liquids e.g. CCl$_{2}$F$_{2}$ (Freon 12) are dispersed in an 
elastic, clear
polymerized gel whose composition has been adjusted to obtain a uniform density,
equal to the liquid freon density (1.3 g/cm$^{3}$). Under ambient 
temperature and pressure the droplets remain in a metastable,
superheated condition. When the critical energy is deposited in a superheated
freon droplet, there is a sudden phase transition during which the droplet 
is vaporized and expands
into a bubble of freon gas of about 1mm in diameter, which is contained 
by the gel. Such an event is
accompanied by an acoustic shock wave, which can be detected with piezoelectric
transducers. Counters which are commercially available have recoil detection 
thresholds as low as several keV to several tens of keV, i.e. precisely in the 
range of sensitivity needed for our application and they are insensitive
to $\beta$- and $\gamma$- radiation. Fig.[1] shows for comparison the mean 
recoil energies which we expect in a Dark Matter search experiment for a 
detector liquid like Freon-12 and another freon type, CF$_{3}$Br, and for 
neutralino masses 
ranging from 10 GeV/c$^{2}$ to 10$^4$ GeV/c$^{2}$.\\
\begin{figure}[h]
\caption{Count rate as a function of operating temperature for Freon detectors
with 3\% loading and doped with 10 Bq of $^{241}Am$ $\alpha$-activity.}
\end{figure}

 The devices used for our tests were obtained
from Bubble Technology Industries \cite{bti}. 8.1 ml of an elastic, 
polyacrylamide based 
polymer is contained in a transparent polycarbonate test tube. The device is 
equipped with a piston. Upon    
unscrewing the piston, the detector becomes sensitive at atmospheric pressure. 
After radiation induced phase transition the 
bubbles are held in place by the gel until the piston is reset and the 
gas bubbles are again reduced to liquid droplets. Properly recompressed, the 
detectors can be used over years. Standard detectors for dosimetry are loaded
with about 0.4\% active material. For our studies we use custom made 
detectors with 3\% loading.\\

\section{Systematic Studies}
\subsection{Sensitivity of Freon-12 to Nuclear Recoils}
 In order to understand the response of these detectors and in particular
in order to get an estimate for the relevant quantities $E_{min}$ and 
$L$, which determine the threshold for Dark Matter detection, 
we doped detectors with 3\% Freon-12 loading with a known 
$\alpha$-activity of $^{241}$Am (10 Bq). The $^{241}$Am was introduced as
a soluble salt and uniformly 
distributed in the polymer, but is not present in the freon itself. 
The measured count rate as a function of operating temperature in the range of 
5$^{0}$C to 45$^{0}$C is shown in fig.[2]. 

Qualitatively the observed 
behaviour can be understood as follows: depending on the respective 
specific energy losses, the detectors are sensitive at low temperatures to
recoil nuclei from Coulomb-scattering of incoming 
$\alpha$'s and to the $^{237}$Np recoils with an energy of 93 keV, which  
follow the emission of 
an $\alpha$- particle of 5.4 MeV. The $^{237}$Np-recoils, however, have a 
range of 
about 100nm only and therefore cannot account for more than 10\% of 
the observed bubble count rate in the temperature range from 20$^{0}$ to 
40$^{0}$C. At 40$^{0}$ we observe a net increase in count rate. Here
the detector is sufficiently sensitive that $\alpha$-particles themselves 
can trigger bubble formation.\\

\begin{figure}[h]
\caption{Energy deposited within the thermal spike length $L$. The curves show
a linear rise as long as the particle range is smaller than $L$ and a flat part
given by $E_{dep} = (dE/dx)L$. Droplet vaporization occurs for recoil energies
with $E_{dep} \ge E_{min}$. At 35$^{0}$C the detector is sensitive only for
recoiling fluorine and chlorine nuclei.}
\end{figure}

For a more quantitative analysis we calculated ranges and stopping powers
for $\alpha$-particles and recoil nuclei in 
Freon with the TRIM program package \cite{trim}. In particular for 
detector nuclei struck 
by an $\alpha$-particle we find
$dE/dx$-values of 1137 keV/$\mu$m, 1074 keV/$\mu$m and 743 keV/$\mu$m for 
chlorine, fluorine, and carbon at their maximum energies of 2.2 MeV, 
3.4 MeV and 
4.5 MeV, respectively. The stopping powers then slowly decrease with decreasing 
energy. For 5.4 MeV $\alpha$-particles the stopping power peaks around 
550 keV with $dE/dx$ = 150 keV/$\mu m$. We assume the same temperature 
dependence for $E_{min}$ as for $E_{c}$ and also the same temperature dependence
for the thermal spike length $L$ as for the critical radius $R_{c}$. 
A 
benchmark point for fixing the scale of $E_{min}$ originates from the 
response of the detectors to 
thermal neutrons following the reaction $^{35}Cl(n,p)^{35}S$ 
(Q-value = 598 keV).
The sulfur recoils with an energy of 17 keV (the proton adds another 1 keV), 
has a short range of 45 nm, a specific energy loss of 390 keV/$\mu$m and
bubble formation is observed at temperatures above 19$^{0}$C 
(\cite{wang}, \cite{harp}).
We then determine 
the deposited energies $E_{dep} = (dE/dx) L$ for recoil nuclei and 
$\alpha$-particles by varying $L$ 
in the interval 
$0.1R_{c} \le L \le 2R_{c}$ and compare 
the resulting $E_{dep}$-values with the minimum energy $E_{min}(T)$ 
needed for bubble formation. 
As shown in tab.[1] a thermal spike length of 
$L_{min} \approx R_{c}$
gives a consistent description of all the data and predicts correctly the 
onset of sensitivity to Cl- and F- recoils between 0$^{0}$ and 5$^{0}$C.
The detector becomes sensitive to C- recoils above 15$^{0}$C and direct
$\alpha$- sensitivity sets in at 40$^{0}$C.\\

\begin{table} 
\begin{tabular}{|l|r|r|r|r|r|r|r|r|r|r|}\hline
T $^{0}$C & 0$^{0}$ & 5$^{0}$ & 10$^{0}$ & 15$^{0}$ & 
20$^{0}$ & 25$^{0}$ & 30$^{0}$ & 35$^{0}$ & 40$^{0}$ \\ \hline 
E$_{c}$[keV]& 6.0 & 3.2 & 2.0 & 1.2 & 0.72 & 0.40 & 0.24 & 0.14 & 
0.08 \\ \hline
R$_{c}$[$\mu$m] & 0.1 & 0.09 & 0.06 & 0.05 & 0.04 & 0.03 & 0.025 & 
0.02 & 0.015 \\ \hline
E$_{min}$[keV]& 150 & 80 & 50 & 30 & 18 & 10 & 6.0 & 3.5 & 2. \\ \hline
E$_{dep}^{\alpha}$[keV] & 15 & 13 & 9 & 8 & 6 & 4.5 & 4 & 3 & 2.25 \\ \hline
E$_{dep}^{Cl}$[keV] & 114 & 97 & 68 & 60 & 45 & 34 & 28 & 23 & 17 \\ \hline
E$_{dep}^{F}$[keV] & 107 & 92 & 64 & 56 & 43 & 33 & 27 & 22 & 16 \\ \hline
E$_{dep}^{C}$[keV] & 74 & 63 & 45 & 39 & 30 & 22 & 18 & 15 & 11 \\ \hline
E$_{dep}^{S}$[keV] & 18 & 18 & 18 & 18 & 16 & 12 & 11 & 8 & 6 \\ \hline
\end{tabular} 
\caption{Temperature dependence of the parameters $E_{c}$, $R_{c}$, $E_{min}$,
which are relevant for droplet vaporization in Freon 12. The critical 
energy and
radius, $E_{c}$ and $R_{c}$ are predicted by the Seitz theory of bubble 
formation ([8],[9]).
$E_{dep} = (dE/dx) L$ is the energy deposited along the thermal spike 
length $L$. $E_{min}$ is the energy which a particle effectively has to 
deposit in the liquid within
a distance $L$ in order to trigger a phase transition. $E_{min}$ and $L$ have
to be inferred from experimental data.}
\end{table}

Using the above results and knowing that the $^{35}$Cl-, $^{19}$F-, 
and $^{12}$C- nuclei
recoil with energies up to 70 keV following Dark Matter particle interaction,
we can determine the respective detection thresholds as a function of 
temperature. Fig.[3] shows the deposited energy $E_{dep}$ as a function of 
the recoil energy $E_{rec}$ for the
three freon constituents at 35$^{0}$C.\\ 

The curves show a linear rise as long as
the particle range is smaller than the thermal spike length $L(T)$ 
and a flat
part given by $E_{dep} = (dE/dx) L(T)$. Since the minimum 
amount of energy 
deposition needed is $E_{min}$ = 3.5 keV at this temperature, the 
detector is only 
sensitive for fluorine and chlorine, with recoil thresholds $E_{th}$ = 7.0 keV 
and 3.5 keV
respectively. The temperature dependence of the detection threshold for 
$^{35}$Cl is summarized in fig.[4]: At 25$^{0}$C a 10 keV chlorine recoil 
can still 
be detected (but not fluorine and carbon recoils!), at 15$^{0}$C the detector is 
completely insensitive to dark matter particles of any mass, but as we saw 
before, the detector remains sensitive to $\alpha$-induced recoils due to 
U/Th- contaminations. This temperature dependence thus provides a means to 
discriminate background contributions.

\begin{figure}[h]
\caption{Temperature dependence of detection threshold for $^{35}Cl$. At 
15$^{0}$C the detector is completely insensitive to neutralinos 
of any mass.}
\end{figure}

\subsection{Droplet and Bubble Diameter Distributions}
 Two groups of detectors of identical composition (in terms of percentages
of freon content or loading, droplet size distributions) were used in order
to study droplet and bubble diameter distributions. 
One group of measurements was performed with our $\alpha$ - emitting 
$^{241}$Am doped detectors. 
The second set of detectors did not include
any radioactive material. At the laboratory temperature (22$^{0}$ C) the 
latter were exposed to an
Ac-Be neutron source, and the bubbles were counted and measured. The same 
measurements were performed 
with the Am-activated bubble detectors. In both cases bubble diameters were 
measured using a calibrated
microscope-video camera-TV monitor  arrangement. Several hundred bubble 
diameters  were measured within 15 min after bubble formation to obtain
bubble diameter distributions. While making the measurements, the detectors 
were monitored by piezoelectric
sensors to ensure that all bubbles were detected and measured. An essentially 
100\% counting efficiency
was achieved so that no apparent distorsion of the distributions by 
observational biases can be expected.\\

\subsubsection{\bf \it Neutron irradiation} 
In the case of the detectors 
irradiated with neutrons, the size 
distributions dN(R)/dR appeared
to be identical for all the detectors studied. In order to understand dN(R)/dR, 
we have to know the droplet
size distribution dn(r)/dr and assume a physical mechanism for transforming the 
neutron energy into an activation energy for triggering the liquid-gas 
phase transition in the freon droplets. We measured dn(r)/dr
for three different detectors of 3\% freon-12 loading and obtained the 
same distribution. Combining the three sets of measurements into a single 
distribution, it was then possible to fit a mathematical
expression with good accuracy. As can be seen on fig.[5a], the droplet 
distribution peaks at 10$\mu$m diameters and tails off at larger 
values. An expression of the form
\begin{equation}
dn(r)/dr = A_{0} (r-p_{1}) [e^{-p_{2}(r-p_{1})^{2}} + p_{3} e^{-(r-p_{1})}]
\end{equation}
appears to describe adequately the observed distribution. As described before, 
at any given temperature, the neutron is detected through a recoiling nucleus
of Cl, F or C within a freon droplet, provided the recoiling nucleus deposits 
a certain minimum amount of energy E$_{min}$ along a path of minimum 
length $L$. Given the value of $L$  
(at 22 $^{0}$C) and the droplet size distribution, it follows 
that the probability for a droplet to undertake a phase transition is 
proportional to its volume (hence to r$^{3}$).
If we further assume that R = $\beta$r, it follows that
\begin{equation}
dN(R)/dR = A_{1}(\beta r)^{3}dn(\beta r)/d(\beta r)
\end{equation}
where only one parameter $\beta$ (apart from an obvious amplitude 
parameter A$_{1}$) is 
left free. It is seen from fig.[5b] that it is thus possible to obtain a 
very good fit to the experimental dN(R)/dR distribution for a value of 
$\beta$ = 8.0, close to the simple approximation $R = 
r [\rho _{l} /\rho _{g}]^{1/3}$, 
where $\rho_{l}$ = 1.3 g/cm$^{3}$ is the liquid freon density  and 
$\rho_{g}$ = 1.25x10$^{-3}$ g/cm$^{3}$ is the freon gas density. 
Assuming these values one gets $\beta$ = 10.\\

\begin{figure}
\caption{a) Droplet size distribution in detectors with 3\% loading. 
b) Distribution of Freon gas bubbles following irradiation with 
fast neutrons. The probability to vaporize is proportional to its volume. 
The hatched areas show droplets which contribute to bubbles with diameters
larger 100$\mu$m. c) Bubble size distribution obtained with detectors 
containing a 10 Bq $^{241}$Am $\alpha$-activity. The main contribution is due 
to Cl, F and C recoils following elastic scattering of $\alpha$-particles.}
\end{figure}

          It is readily seen from the distributions that although 50\% of 
the droplets have diameters smaller than 13 $\mu$m, these droplets 
contribute only to 1-2 \% of the bubbles produced in the irradiation of
our detectors by neutrons. The hatched areas on fig.[5a] and fig.[5b] show the 
droplets which contribute to the production
of bubbles of diameters larger than 100 $\mu$m. In other words, given 
the energy loss values for various recoiling nuclei, and the size of the 
thermal spike length $L$, the neutrons essentially 
sample the volume distribution
of the droplets, and there are not enough small droplets to make a 
significant contribution to the neutron detection efficiency of our bubble 
detectors.\\

\subsubsection{\bf \it Response to $^{241}$Am $\alpha$-particles} 
 A similar analysis was made for the bubble size distributions obtained with 
the detectors that contained an $^{241}$Am activity. For these detectors, the 
droplet distribution was the same as for the neutron detectors. At the 
temperature of 22$^{0}$C at which the measurements were performed, 
the detectors were 
insensitive to $\alpha$-particles, but sensitive to Cl, F, C recoils 
following Coulomb scattering with $\alpha$-particles and to the 
$^{237}$Np recoils from the $^{241}$Am decay itself. Because of the very 
short range of the $^{237}$Np recoils, only 
$\alpha$-decays occurring in a very thin shell near the surface of the 
droplets can trigger bubble formation, and the bubble-size distribution 
expected for this mechanism would be of the form 
$ dN(R)/dR = A_{2}(\beta r)^{2} dn(\beta r)/d(\beta r)$, but due to the 
short range of the 
recoils this "surface" contribution has
a negligible effect on the overall bubble diameter distribution.\\

\begin{figure}[h]
\caption{Explosive droplet vaporization is accompanied by an acoustic
signal which can be recorded by piezoelectric transducers. All bubble 
signals start off with a sharp negative spike lasting about 2$\mu$sec.}
\end{figure}

The $^{35}$Cl-, $^{19}$F-, $^{12}$C- recoils give the main contribution 
to droplet evaporation at 22$^{0}$C. Given the $\alpha$-particle range
in the gel or liquid freon (50-60 $\mu$m), at the (relatively) low 
temperature used for the measurements, two
different bubble size regimes can be identified. For droplet sizes much 
smaller than the $\alpha$-range (remember that
50\% of the droplets have diameters smaller than 13$\mu$m), the 
$\alpha$-particles essentially reach through the entire
droplet volume. One should therefore expect a $R^{3}dn(R)/dR$ form to yield a 
good description of the bubble distribution for 
small diameter values. For larger R-values, on the other hand, at 
relatively low temperatures (and large critical energy values),
only the more energetic recoils will trigger bubble formation, so that 
one expects the active recoils to be concentrated
in a more or less thick shell near the droplet surface. This is well 
illustrated in fig.[5c] where an $R^{2}dn(R)/dR$ expression best fits 
the bubble 
size distribution above the maximum, and an $R^{3}dn(R)dR$ curve does better 
for smaller
diameter values. Here again, the small droplet contribution is strongly 
suppressed because of the relatively long
range of the $\alpha$-particles.

\subsection{Piezoelectric Signal Detection and Event Localization}
 In detectors with more than 3\% loading the elastic medium 
becomes opalescent, non-transparent and eventually optical detection of the 
bubbles becomes impossible. For this reason and also in order to be able 
to trigger electronically on the events we investigated acoustic event 
detection of bubble formation. Piezoelectric sensors had already been 
shown to be sensitive to the sound created by a vaporizing droplet 
\cite{apf}. In order to simulate a larger detector volume several detector
test tubes were immersed in a (28 x 28 x 28) cm$^{3}$ water tank, with four 
piezoelectric sensors installed at the tank walls in such a way that the 
transducer surface 
made direct contact with the water. We use wide band sensors 
(CANDED type-WD) 
with a sensitivity bandwidth ranging from 100 to 1000 kHz. Their 
signals are fed
into low noise current sensitive preamplifiers LM6365 with a gain of 
15V$/$$\mu$A and a noise characteristic of 1.5 $pA/\sqrt{Hz}$. 
The signals from bubble formation are very characteristic (fig.[6]) and easily 
distinguishible from background signals: all bubble signals start off 
with a sharp negative spike lasting about $2\mu$sec; after this, 
strong oscillations at about 25 kHz arise and the amplitudes of these 
oscillations are then modulated with much slower (kHz) frequencies, 
which vary from event to event and are probably due to sound reflections in the 
container.\\
\begin{figure}
\caption{Pulseheight distributions recorded in a coincidence experiment
with two piezoelectric sensors. A close sensor is mounted at a fixed distance
2cm from a detector test tube and a far detector is moved up to 20cm apart.
The amplifier noise is 3mV (rms).}
\end{figure}

 Fig.[7] shows the pulse height distributions recorded in a 
coincidence
experiment with two sensors, a close sensor at a fixed distance of 2 cm from a
detector test tube and a far sensor at distances varying from 2 cm 
up to 20 cm. At the 
far 20 cm position the signals are still clearly separated from the 3 mV rms 
noise band and we find 100\% efficiency to detect bubble formation. A video 
recording allowed correlating the electronic signals to individual bubbles in 
the (low loading) detectors; under these condirions full agreement was found, 
i.e. 
whenever a bubble was detectable visually, also an electronic signal was 
observed and vice versa. 

From this, together with the observed bubble size 
distribution discussed in the previous section, we can infer a lower cut-off
in the droplet size distribution at around 15$\mu$m. Since most of the 
detector mass is concentrated in larger size droplets, this cut-off translates 
into a freon volume efficiency of 95\%.\\ 

Our tests furthermore showed that acoustic time-of flight methods can be 
used to locate 
bubbles. With four sensors the signals were fed 
into a FADC system and the
event position is calculated using a neural network trained for our detector
geometry.  These tests indicate that an event localization is possible 
with a resolution of 0.5 cm in the tank.\\

\subsection{Background Considerations}
Being not directly sensitive to $\alpha, 
\beta$ and $\gamma$ - radiation is the big asset of our detector. Still
there are several sources of background, internal and external to the detector
which we have to consider:
\begin{itemize}
\item The superheated droplets will be sensitive to the 100 keV nuclear 
recoils following the 5 to 6 MeV $\alpha$- decays due to the presence of U/Th 
daughters in the detector material, i.e. the freon itself. It is therefore
mandatory to  bring the contamination of $^{235}$U 
and $^{232}$Th down to the level of 10$^{-15}$g/g in order 
to reach a sensitivity of the order 10$^{-3}$cts/kg/d. Since Freon-12 
is a cryogenic liquid, we are confident that such a low level of 
contamination is attainable.
\item The detector will register with high efficiency recoiling nuclei 
in  droplets hit by
$\alpha$-particles from $^{235}$U and $^{232}$Th decays in the polymer.
An estimate for the tolerable contamination in the polymer can be obtained
from the observed countrate of 0.05 cts/s in our 10 Bq $^{241}$Am spiked 
detectors. From that we conclude that we need a U$/$Th purity of 
2x10$^{-13}$g/g in the 
polymer. Since the elastic gel consists to a large part ($\approx$ 80\%) 
of water
which is available with a purity better than 5x10$^{-15}$g/g U/Th \cite{sno},
a purity of the rest of the material at the order of 1x10$^{-12}$ g/g is 
acceptable.
\item With the detector installed in a deep underground laboratory (e.g. SNO)
cosmic muon induced background is reduced to a negligible level but
consideration has still to be given to the fast neutron component coming
from the rock walls. With the measured flux of 3 x 10$^{-6}$ n/cm$^{2}$/s
a passive shielding of 1m (borated) water will reduce the neutron
induced countrate to a level of several 10$^{-3}$n/d.
\item Spontaneous droplet vaporization is another potential background 
source, which has still to be addressed. Manufacturers of 
droplet detectors and \cite{col} claim that this background is essentially
non-existent, but no systematic studies at the level of our sensitivity exist
at the moment.

\end{itemize}

\begin{figure}
\caption{ Neutralino induced recoil spectra for coherent scattering 
on $^{35}$Cl. The assumption is made that the dark matter particles follow a
Maxwellian velocity distribution in the galactic halo. For the cross 
section we assume coherent interaction via Higgs boson exchange, 
as described e.g. in [1].}
\end{figure}

In order to assay the potential intrinsic background and to understand its 
origin, measurements were performed on the surface and at the location 
of the Sudbury Neutrino Observatory (SNO) 2000 m under ground with and 
without 20 cm 
water shielding. At the surface 0.5 cts/d were 
recorded for standard detectors with 0.4\% loading; in the mine two detectors 
were exposed up to now and showed no counts after 80 days. The detector masses 
involved in these 
measurements were however too small to allow meaningful conclusions about the
intrinsic background of the devices. Independent of these direct measurements,
the detector material was also tested for radioactive contaminations at the 
Gran Sasso low activity counting facility (\cite{christ}). Limits of 2.2 Bq/kg
were obtained for the $^{232}$Th contamination and again the measurements were 
compromised by the small mass of the test samples. On the other hand  
substantial $^{134}$Cs, $^{137}$Cs activities of about 0.2Bq/kg were found 
in the samples (this is due 
to the presence of CsCl salt, which is mixed into the gel in order to match 
the gel density to the  density of liquid Freon). Although the associated
$\beta$- and $\gamma$ activities are not harmful by themselves for our 
application, we suspect an
important contamination of U/Th of the unpurified salt itself. This point 
is subject to further clarification.

\section{Detector Sensitivity and Achievable Limits for Neutralino Detection}
The relevant quantity which has to be determined in order to compare 
theoretical expectations to experimental data
is the nuclear recoil spectrum following elastic scattering of a dark matter 
particle with a freon atom and the effect of the energy threshold, which in 
turn depends on the operating temperature of the detector. If we assume that the
dark matter particles are distributed in our galaxy in a self gravitating halo
with an isotropic Maxwellian velocity distribution, one can show that the   
recoil spectrum can be well approximated by an exponentially falling
distribution of the form
\begin{equation}
dR/dE \approx\,(R_{0}/<E_{R}>) exp (-E/<E_{R}>) F^{2}(E_{R})
\end{equation}
$<E_{R}>$ is the mean recoil energy already described in eq.(1) and $F(E_{R})$ 
is a form factor, which describes in the case of coherent neutralino-nucleus 
interactions the coherence loss for non-zero momentum 
transfer due to the finite size of the nucleus; for light nuclei 
like $^{35}$Cl it 
is close to one (\cite{gould}). R$_{0}$ is the total rate in cts/kg/d
\begin{equation}
R_{0} = N_T (\rho_X/M_X) \sigma <v>
\end{equation}
where $N_{T}$ denotes the number of target atoms, $\rho_{X}/M_{X}$ is the 
number density of neutralinos, $\sigma$ the neutralino interaction cross 
section and $<v>$ the relative average velocity.

For an evaluation of the 
cross section $\sigma$ in the Minimal Supersymmetric Standard 
Model (MSSM) we follow the treatment in ref.\cite{bot}. The experimentally most favorable
situation occurs for a coherent interaction with the whole nucleus via 
Higgs boson exchange, provided the neutralino is 
a zino-higgsino mixture. Setting the MSSM parameters to $\mu > 0$, 
tan$\beta$ = 8 and with a Higgs mass of 50 GeV/c$^{2}$ one obtains an isotropic 
elastic cross section of the order of
\begin{equation}
\sigma_{coh} = 4.25\,10^{-40}A^{2}(M_{N}M_{X})^{2}/(M_{N}+M_{X})^2\, [cm^{2}] 
\end{equation}
where A is the mass number of the nucleus. If the zino-higgsino mixing is small
the coherent cross section is strongly suppressed by one or two orders of 
magnitude and spin dependent interactions might give more important 
contributions. For the coherent case practically only the neutralino 
interactions with chlorine matter, while for the spin dependent case, as for 
photinos, the 
$^{19}$F isotope with spin 1/2$^{+}$ 
gives the biggest contribution (\cite{ellis}). Fig.[8] shows the recoil
spectra for $M_{X}$ = 10, 10$^{2}$ and 10$^{3}$ GeV/c$^{2}$ for coherent 
scattering on$^{35}$Cl.\\

\begin{figure}
\caption{ Detection efficiency as function of neutralino mass. The effect 
of the temperature dependent energy thresholds on the differential recoil
spectra of chlorine is apparent.}
\end{figure}

\begin{figure}
\caption{The total count rate for coherent scattering on $^{35}$Cl shows a
temperature dependent cut-off at small masses. The broken line represents 
the interaction rate without these detector effects.}
\end{figure}

 In order to calculate total count rates and sensitivities, the 
effect of the temperature - dependent energy threshold on the differential 
recoil spectra has to be included. The resulting detection efficiency as 
a function of the
neutralino mass is shown in fig.[9] and as is evident from the recoil spectra,
smaller masses are more strongly affected. The total count rate for coherent 
scattering on $^{35}$Cl in fig.[10] shows a temperature dependent cut-off 
at small 
masses below M$_{X}$=10 GeV/c$^{2}$. The interaction rate peaks where the 
neutralino mass matches 
the mass of the target nucleus and it decreases with increasing 
neutralino mass. In Fig.[11] we show the behaviour of the upper
limits on the cross section for a detector with a background at the level of 
10$^{-3}$ cts/kg/d.
 
\begin{figure}[h]
\caption{Curve a) shows the upper limit on the neutralino cross section 
which can be set by
a superheated droplet detector with a background at the level of 
10$^{-3}$cts/kg/d. The  weak cross section for a heavy Dirac neutrino is
indicated by curve b).  For
comparison curves c) and d) give experimental limits obtained with 
Ge-detectors. 
c) is the exclusion plot for weakly interacting massive particles obtained by
the Neuchatel/Caltech/PSI Ge-experiment in the St. Gotthard tunnel 
and d) is the result of the Heidelberg-Moscow Ge-experiment in the 
Gran-Sasso Laboratory [21].}
\end{figure}

\section{Conclusions and Outlook}
 
  We discussed the use of moderately superheated liquids in the form of
superheated droplet detectors for a new type of dark matter search experiment.
The advantage of this method for Dark Matter detection is that the
detector material is cheap, readily available and that it is easily possible 
to fabricate a large mass detector. From its easy operating conditions and 
its suitable isotopic composition ($^{19}F$ is a spin-$1/2$$^{+}$ isotope) 
CCl$_{2}$F$_{2}$ (i.e. Freon 12) is an interesting active material. Even more 
attractive because of its higher mass could be CF$_{3}$Br. It has a similar 
vapor 
pressure curve as Freon-12 and is also non-flammable. Compared to 
alternative  
techniques our technique is insensitive to $\alpha$, $\beta$ and $\gamma$ 
radiation, and avoids the need of cryogenics. However the detector is sensitive
to neutron and $\alpha$- particle induced recoils. Therefore care has to be 
taken to shield away neutrons and to reduce U/Th cotaminations in the Freon 12 
itself.\\

An evaluation of the various parameters of our detector system suggests a 
modular approach. In particular we are considering a unit based on a 
cylindrical emulsion volume of 10 kg emulsion. With an achievable loading 
of 10\% we thus obtain 
a 1kg active
detector module, read with an array of 6 piezoelectric sensors 
connected to a data
acquisition system. Such a detector will represent a realistic prototype of
a droplet based detector for more detailed background studies on surface 
and underground. At the same time it would serve as a modular unit easily 
augmented to obtain the large-mass system envisaged for the search for 
weakly interacting massive particles.\\

{\bf Acknowledgements}\\

We are indepted to H. Ing and R. Noulty from BTI Bubble Technology Industries 
for many helpful discussions and for providing us with various samples of
specially developed superheated droplet detectors. We wish to thank 
C. Hargrove, Carleton University, for fruitful discussions on purification 
techniques. We acknowledge precious help from our collegues J.P. Martin and
M. Beaulieu in data acquisition and electronics issues. This work was 
supported in part by the National Research and Science Council of Canada.


\newpage

\begin{thebibliography}{99}
\bibitem{bot} A. Bottino et al., Phys. Lett. B295 (1992) 330
\bibitem{griest} K. Griest et al., Phys. Rev. D41 (1990) 3565
\bibitem{mcdo} J. McDonald et al., Phys. Lett. B283 (1992) 80
\bibitem{good} M.W. Goodman and E. Witten, Phys. Rev. D41 (1990) 2388
\bibitem{zac1} V. Zacek, Il Nuovo Cimento, 107A (1994) 291
\bibitem{zac2} V. Zacek, Proceedings of the 2$^{nd}$ Workshop on "The
  Dark Side of the Universe" Universita di Roma II, November 1995, Rome, Italy
\bibitem{col} J.I. Collar, Phys. Rev. D54 (1996) 1247
\bibitem{seitz} F. Seitz, Phys. Fluids 1 (1958) 2
\bibitem{wang} W. Lim and C.K. Wang, Nucl. Inst. and Meth. A 336 (1993) 215
\bibitem{harp} M.J. Harper and J.C. Rich, Nucl.Inst. and Meth. A336 (1993) 220
\bibitem{riep} G. Riepe and B. Hahn, Helv. phys. Acta, 34 (1961) 865
\bibitem{apf} R. Apfel, Nucl. Inst. and Meth. 162 (1979) 603
\bibitem{ing} H. Ing and B.C. Bimboim, Nucl. Tracks 8 (1984) 285
\bibitem{saw} J. Sawicki, Nucl. Inst. and Meth. A 336 (1993) 215
\bibitem{bti} BTI, Bubble Technology Industries, Chalk River, Ontario, Canada
\bibitem{trim} J.F. Ziegler, TRIM Version 95.4; J.F. Ziegler, J.P. Biersack and 
U. Littwark "The stopping Powers and Ranges of Ions in Solids", 
N.Y. Pergamon Press (1985) 
\bibitem{sno} The SNO-collaboration, private communication
\bibitem{christ} We thank the LNGS low background counting team for their help
\bibitem{gould} A. Gould, Astr. Phys. Journ. 321 (1987) 571
\bibitem{ellis} J. Ellis and R. Flores, Phys. Lett. B263 (1991) 259
\bibitem{beck} M. Beck, Proceedings of the Workshop on "The
  Dark Side of the Universe" Universita di Roma II, June 1993, Rome, Italy
\end{thebibliography}
\end{document}